\documentclass[a4paper,nofootinbib,showpacs,preprintnumbers,floatfix,superscriptaddress,pra,twocolumn]{revtex4}
\usepackage{amsmath, amsthm, amssymb}
\usepackage{graphicx}
\usepackage{dcolumn}
\usepackage{bm}
\usepackage{color}
\usepackage{amsmath}
\usepackage{amsfonts}
\usepackage{amssymb}

\newcommand{\ket}[1]{| #1 \rangle}
\newcommand{\bra}[1]{\langle #1 |}

\newcommand{\eg}{{\it{e.g.~}}}

\newcommand{\beq}{\begin{equation}}
\newcommand{\eeq}{\end{equation}}
\newcommand{\bea}[1]{\begin{equation}\begin{array}{#1}}
\newcommand{\eea}{\end{array}\end{equation}}
\newcommand{\beqn}{\begin{eqnarray}}
\newcommand{\eeqn}{\end{eqnarray}}
\providecommand{\openone}{\mathbbm{1}}
\renewcommand{\rho}{\varrho}

\begin{document}
\title{Robust multipartite quantum correlations without complex encodings}
\author{Rafael Chaves}
\affiliation{ICFO-Institut de Ciencies Fotoniques, Mediterranean
Technology Park, 08860 Castelldefels (Barcelona), Spain}
\author{Leandro Aolita}
\affiliation{ICFO-Institut de Ciencies Fotoniques, Mediterranean
Technology Park, 08860 Castelldefels (Barcelona), Spain}
\author{Antonio Ac\'in}
\affiliation{ICFO-Institut de Ciencies Fotoniques, Mediterranean
Technology Park, 08860 Castelldefels (Barcelona), Spain}
 \affiliation{ICREA-Instituci\'o Catalana de Recerca i
Estudis Avan\c cats, Lluis Companys 23, 08010 Barcelona, Spain}

\begin{abstract}
One of the main challenges for the manipulation and storage of multipartite entanglement is its fragility under noise. We present a simple recipe for the systematic enhancement of the resistance of multipartite entanglement against any local noise with a privileged direction in the Bloch sphere. For the case of exact local dephasing along any given basis, and for all noise strengths, our prescription grants full robustness: Even states with exponentially decaying entanglement are mapped to states whose entanglement is constant. In contrast to previous techniques resorting to complex logical-qubit encodings, such enhancement is attained simply by performing local-unitary rotations before the noise acts. The scheme is therefore highly experimentally friendly, as it brings no overhead of extra physical qubits to encode logical ones. In addition, we show that, apart from entanglement, the resiliences of the relative entropy of quantumness and the usefulness as resources for practical tasks such as metrology and nonlocality-based protocols are equivalently enhanced.
\end{abstract}

 \maketitle
\emph{Introduction.---} Multipartite quantum correlations in
composite systems subject to local noise are in general extremely
fragile, typically decaying more quickly as the number $N$
of particles increases. For instance, one of the most important genuinely
multipartite entangled states is the Greenberger-Horne-Zeilinger
(GHZ) state \cite{GHZ}. It represents coherent superpositions of
the kind of the celebrated Schr\"odinger's cat state~\cite{cat}.
Unfortunately, under the action of local noise, its entanglement
decays exponentially fast, with a decay-rate that grows
proportionally to $N$~\cite{expdecay,expdecayrobustdecay,aolita}.
Thus, the system is very rapidly taken into, or close to, a
separable mixture. This exponential fragility entails serious
drawbacks for the practical applicability of GHZ states as
resources for quantum information processing in realistic
scenarios. For example, for any noise strength, the quantum gain
provided by GHZ entanglement in parameter
estimation~\cite{metrology} or distributed-computing
scenarios~\cite{CCP,CCPBrukner,chaves} vanishes almost
instantaneously already for a moderate system size.

A possible way to enhance the robustness of quantum correlations
is to encode logical qubits into error-correction codewords,
consisting of entangled states of many physical
qubits~\cite{expdecayrobustdecay,dur}. For instance, for small
noise strengths $p$, the decay rate of logical GHZ entanglement
under local white noise can be decreased exponentially with the
number of physical qubits in the codeword when the codewords are
themselves GHZ states~\cite{dur}. This is remarkable because the
enhancement is achieved passively, {\it i.e.} without any active
error correction. However, there is a price to pay in experimental
overhead: For the logical state to achieve full entanglement
robustness -- in the sense that its logical entanglement becomes
independent on $N$--, each logical qubit requires a number of
genuine-multipartite entangled physical qubits that scales
logarithmically with the number of logical qubits.

With the maximally mixed state as its only steady state, local
white noise (local depolarization) is the most detrimental type of
local noise. Nevertheless, in many realistic situations the noise
can be assumed, up to good approximation, to possess privileged
directions in the Bloch sphere, including pure states as steady
states. This is the case, for instance, in many experiments with
atomic or ionic qubits, where the dominant source of noise is
dephasing from magnetic-field and laser-intensity fluctuations,
and from spontaneous emissions during Raman couplings
\cite{ion_review}. Another example is provided by birefringent
polarization-mantaining optical fibers
\cite{pol_mantaining_fibers}, where mechanical stress and
temperature induce index-refraction fluctuations that dephase
polarization qubits. In the former case, the privileged noise
direction is that of the quantization axis defined by the magnetic
field, while in the latter, that of the linear polarizations along
the ordinary and extraordinary axes of the fiber.

In this work we study the action  on graph states of local noisy
channels with an approximately well-defined privileged basis.
Graph states constitute a family of genuine multi-qubit entangled
states with remarkable applications~\cite{graph_review}. Relevant
examples thereof are the previously mentioned GHZ state, or the
cluster state, which allows for measurement-based quantum
computation~\cite{cluster}. We introduce an experimentally
friendly recipe, consisting of local-unitary rotations before the
noise acts, to enhance the resistance of graph-state quantum
correlations. Remarkably, and despite its simplicity, for exact
dephasing this prescription supplies the states with full
robustness: It gives an $N$-independent lower bound for the decay
of graph-state entanglement. In particular, the exponentially
fragile entanglement of GHZ states is enhanced to decay only
linearly with $p$, for all $N$. In addition, the bound holds for
quantum correlations other than entanglement \cite{Kavan} and is
robust against mixedness in the initial states. Finally, for GHZ
states, we  show that the local-unitary protection resists small noise
deviations from exact dephasing, and that the enhancement applies
also to the usefulness for physical tasks such as metrology
~\cite{metrology} and distributed-computing~\cite{CCP,CCPBrukner}
protocols.

\emph{Enhancement of the robustness of graph-state quantum
correlations.---}\label{sec:robust_entanglement} We consider local
completely positive trace-preserving channels $\mathcal{E}$,
defined on any state $\varrho$ as
\begin{eqnarray}
\nonumber \mathcal{E}(\varrho)&\doteq&(1-\frac{p}{2})
\varrho+\frac{p}{2}( \alpha_{X} X\varrho X \\
&+& \alpha_{Y} Y\varrho Y + \alpha_{Z} Z\varrho Z),
\label{mapdeviation}
\end{eqnarray}
where $X$, $Y$, and $Z$ are, respectively, the first, second, and
third Pauli matrices in the computational basis $\{\ket{0},
\ket{1}\}$. Parameters $0\leq\alpha_{X},\alpha_{Y},\alpha_{Z}\leq
1$ satisfy the normalization condition
$\alpha_{X}+\alpha_{Y}+\alpha_{Z}=1$. The composite $N$-qubit map
$\Lambda$  is given by the single-qubit map composition
$\Lambda(\rho)\doteq \mathcal{E}_{1}\otimes\mathcal{E}_{2}\otimes
...\ \mathcal{E}_{N}(\rho)$, where $\mathcal{E}_k$, with $1\leq
k\leq N$, corresponds to map \eqref{mapdeviation} acting on the
$k$-th qubit. The focus of our attention throughout is in the
situations where $\alpha_{Z}>>\alpha_{X}$, $\alpha_{Y}$, so that
$\Lambda$ is close to local phase-damping map $\Lambda^{PD}$,
which corresponds to $\alpha_{Z}=1$. Probability $0\leq p\leq 1$
measures the noise strength and gives also a convenient
parametrization of time: $p = 0$ refers to the initial time $t=0$
and $p= 1$ refers to the asymptotic $t\rightarrow\infty$ limit.
Note that the $1/2$ factors in \eqref{mapdeviation} are such that
an exact fully-dephasing channel appears at $p=1$ and
$\alpha_z=1$.


Let us begin by the phase-damping channel $\alpha_{Z}=1$. We focus
first on GHZ states
\begin{equation}
\vert {{\Phi_+}^{N}}\rangle\doteq\frac{1}{\sqrt{2}}\big(\left\vert
0\right\rangle ^{\otimes N} +\left\vert 1\right\rangle ^{\otimes
N}\big). \label{ghzstate}
\end{equation}
Under $\Lambda^{PD}$, all the entanglement in \eqref{ghzstate}
decays (at slowest) exponentially with $N$, as
$(1-p)^N$~\cite{aolita}. We show next that, for a fixed $p$, the
entanglement of
\begin{equation}
\ket{{{\Phi_+}^{N}_{T}}}\doteq H^{\otimes N}
\ket{{{\Phi_+}^{N}}}=\frac{1}{\sqrt{2}}\big(\left\vert
+\right\rangle ^{\otimes N} +\left\vert -\right\rangle ^{\otimes
N}\big),
\label{ghzstatetrans}
\end{equation}
under $\Lambda^{PD}$ is independent on $N$. Operator $H$ stands
for the Hadamard-gate rotation, defined by $H\ket{0}\doteq\ket{+}$
and $H\ket{1}\doteq\ket{-}$, with
$\ket{\pm}\doteq\frac{1}{\sqrt{2}}(\ket{0}\pm\ket{1})$.
Transversal states \eqref{ghzstatetrans} are thus local-unitarily
equivalent to \eqref{ghzstate}, possessing therefore the same
amount and type of entanglement.

We consider here an arbitrary entanglement monotone $E$, {\it i.e.} any function of $\varrho$ which is non-increasing under local
operations and classical communication. In Appendix A however, we extend the treatment to the relative
entropy of quantumness  \cite{Kavan}, which is not an entanglement
monotone. Notice first that a single-qubit $Z$ measurement on
$\ket{{\Phi_+}^{N}_{T}}$ leaves the system in
state $\ket{{\Phi_+}^{N-1}_{T}}\otimes\ket{0}$ or $\ket{{{\Phi_-}^{N-1}_{T}}}\otimes\ket{1}$, with
$\ket{{{\Phi_-}^{N-1}_{T}}}\doteq\ket{+}^{\otimes
N}-\ket{-}^{\otimes N}$. Similarly, since it commutes with
$\Lambda^{PD}$, a $Z$ measurement on ${\rho_+}_{T}^{N}\doteq\Lambda^{PD}(\ket{{{\Phi_+}^{N}_{T}}})$ leaves the system in ${\rho_+}_{T}^{N-1}\otimes\ket{0}\bra{0}$, or ${\rho_-}_{T}^{N-1}\otimes\ket{1}\bra{1}$, with ${\rho_-}_{T}^{N-1}\doteq\Lambda^{PD}(\ket{{{\Phi_-}^{N-1}_{T}}})$.
Furthermore, it is immediate to see that ${\rho^+}_{T}^{N-1}\otimes\ket{0}\bra{0}$ and
${\rho_-}_{T}^{N-1}\otimes\ket{1}\bra{1}$ are local-unitarily equivalent. From
this, and the monotonicity of $E$ under local measurements, it
follows then that $E({\rho_+}_{T}^{N})\geq E({\rho_+}_{T}^{N-1}\otimes\ket{0}\bra{0})$.
Iterating this reasoning $N-2$ times and, for ease of notation, omitting the tensor-product factors, one obtains that
\begin{eqnarray}
\label{ordering} \nonumber
E\big(\Lambda^{PD}(\ket{{{\Phi_+}^{N}_{T}}})\big)&\geq& E\big(\Lambda^{PD}(\ket{{{\Phi_+}^{N-1}_{T}}})\big)\geq\\
\hdots&\geq& E\big(\Lambda^{PD}(\ket{{{\Phi_+}^{2}_{T}}})\big).
\end{eqnarray}
That is, transversal states \eqref{ghzstatetrans} possess at least as much
resistance as the two-qubit state $\ket{{{\Phi_+}^{2}_{T}}}$, for all
$N\geq2$. As an example, consider the robustness of $N$-party distillable GHZ-entanglement. The distillation of maximally entangled pairs between any pair of particles is  sufficient to distill an $N$-qubit GHZ state \cite{expdecayrobustdecay}. Bound \eqref{ordering}, for $E$ the $N$-party distillable two-qubit entanglement between any pair, implies that the distillation of entangled pairs (and consequently  also of GHZ states) from the transversal states is at least as robust for $N$ qubits as for $2$, in contrast to bare states \eqref{ghzstate}. In Appendix B, we show that \eqref{ordering} holds not only for GHZ states but actually for arbitrary graph states, and even initially in the presence of
global white noise. That is, the entanglement decay of generic graph states, encoded in appropriate transversal local bases, is bounded from below
by that of a two-qubit graph state. The bound follows again from the fact that any
connected $N$-qubit graph state can be mapped into one of $N-1$ qubits by single-qubit $Z$ or $X$ measurements. The encoding is again given by single-qubit Hadamard rotations, but applied only to the qubits  measured in $X$ in the mapping (see Appendix for details).

\begin{figure} [!t]
\centering
\includegraphics[width=0.8\linewidth]{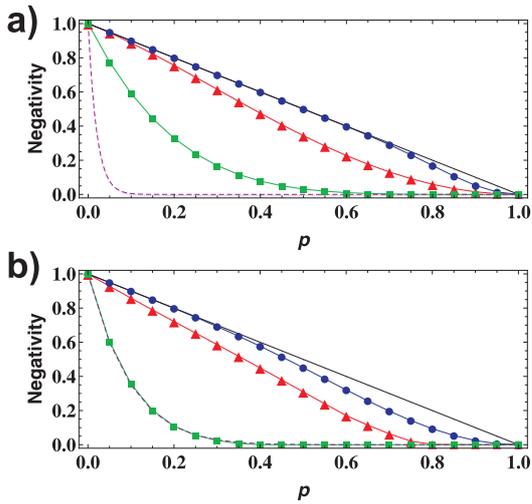}
\caption { (Color online) {\bf a.} Negativities of  bipartitions
``one qubit versus the rest", as a function of $p$, for different
initial pure states under  local dephasing. Already for $N=5$,
the transversal GHZ state $\ket{{{\Phi_+}^{5}_{T}}}$ (red triangles)
displays a considerably higher  robustness than the bare GHZ state
$\ket{{{\Phi_+}^{5}}}$ (green squares). As $N$ goes to infinity,
the negativity corresponding to $\ket{{{\Phi_+}^{N}_{T}}}$ tends
to $1-p$ (thin black line). However, for any finite $N$ there is
always a $p$ up to which this negativity is approximately equal to
the $N$-independent $1-p$ curve; and the smaller the $N$, the
smaller is the $p$ for which this happens. For $N=50$ (blue circles)
for instance, this happens for $p\approx0.65$. The purple dashed
line corresponds to $\ket{{{\Phi_+}^{50}}}$. {\bf b.} Same
negativities, for $N=10$, but with a deviation $\epsilon=0.05$
from exact dephasing. Transversal GHZ states (red triangles) are
still  exponentially more robust than  bare GHZ states (green
squares). The negativity of transversal GHZ states for the case
$\epsilon=0$ (blue circles) is always on top, while the one of
bare GHZ states for $\epsilon=0$ (purple dashed) is almost
identical to that of bare GHZ states for $\epsilon=0.05$. For
comparison, the linear decay $1-p$ is also shown (thin solid
black)}. \label{fig:entanglementinbipartitions}
\end{figure}

\par Next, we probe bound
\eqref{ordering} with a simple-to-calculate entanglement monotone:
the negativity~\cite{negativity}. In addition, to test their
robustness against deviations from exact dephasing, we allow for
maps $\Lambda$ with arbitrary $\alpha_x,\alpha_y$, and $\alpha_z$.
Specifically, we calculate analytically the negativity
$\mathcal{N}$ of any bipartition ``one qubit versus the rest" of
$\Lambda(\ket{{{\Phi_+}^{N}_{T}}})$. The smaller $\mathcal{N}$ is,
the closer is the state to featuring no genuine $N$-qubit
entanglement~\cite{expdecay,expdecayrobustdecay,aolita}. Due to
the GHZ-diagonal structure (see, for instance, Ref.~\cite{aolita}) of
$\Lambda(\ket{{{\Phi_+}^{N}_{T}}})$, the calculation is enormously
simplified, and reduces essentially to diagonalizing $2^{N-1}$
matrices of dimensions $2\times2$. One obtains
\begin{eqnarray}
\label{neg} \nonumber
\mathcal{N}&=& \sum\limits_{\mu=0}^{\lfloor\frac{N-1}{2}\rfloor} \binom{N-1}{\mu}\Big(\max[0,f^{-}_{\mu+1}  -f^{+}_{\mu}]\\
&+&\max[0,f^{-}_{\mu}  -f^{+}_{\mu+1}]\Big),
\end{eqnarray}
where  $f^{\pm}_{\mu}\doteq(\frac{p}{2}\alpha_{Z}\pm
\frac{p}{2}\alpha_{Y})^{\mu}\big( (1-\frac{p}{2})\pm
\frac{p}{2}\alpha_{X}\big) ^{N-\mu}+(\frac{p}{2}\alpha_{Z}\pm
\frac{p}{2}\alpha_{Y})^{N-\mu}\big( (1-\frac{p}{2})\pm
\frac{p}{2}\alpha_{X}\big)^{\mu}$ and
$\lfloor\frac{N-1}{2}\rfloor\doteq N/2-1$, for $N$ even, or
$\lfloor\frac{N-1}{2}\rfloor\doteq (N-1)/2$, for $N$ odd.

For the particular case $\alpha_{Z}=1$ addressed by
\eqref{ordering}, the negativity \eqref{neg} simplifies (see Appendix C) to
\begin{eqnarray}
\label{negZ} \nonumber
\mathcal{N}_{(\alpha_{Z}=1)}&=&(1-p)\sum_{\mu=0}^{\lfloor\frac{N'}{2}\rfloor} \binom{N'}{\mu}\bigg[\frac{p}{2}^{\mu}(1-\frac{p}{2})^{N'-\mu}\\
&-&\frac{p}{2}^{N'-\mu}(1-\frac{p}{2})^{\mu}\bigg],
\end{eqnarray}
with $N'\doteq N-1$. In Appendix C we show that
$\mathcal{N}_{(\alpha_{Z}=1)}\to 1-p$ as $N\to\infty$, for all
$0\leq p<1$. In addition, for all $0\leq p\leq1$, we observe that
the bound on $\mathcal{N}_{(\alpha_{Z}=1)}$ given
by~\eqref{ordering} is too conservative, as it is not tight.
Furthermore, the limit $1-p$ is approached the faster the smaller
$p$. This can be appreciated in Fig.
\ref{fig:entanglementinbipartitions} a), where
$\mathcal{N}_{(\alpha_{Z}=1)}$ for states \eqref{ghzstate} and
\eqref{ghzstatetrans} under the local-dephasing map $\Lambda^{PD}$ is
plotted as a function of $p$ and for different $N$.


In turn, for the generic case $\alpha_{Z}=1-\epsilon\leq 1$, with
$\epsilon\doteq\alpha_{X}+\alpha_{Y}$ the deviation from exact
dephasing, we focus on the weak-noise regime. For sufficiently
small $p$, one can ignore all terms not linear in $p$ and
approximate negativity \eqref{neg} as
$\mathcal{N}\approx1-p(N\epsilon+1-\epsilon)\approx(1-p)^{N\epsilon+1-\epsilon}$.
While this is no longer an $N$-independent behavior, it clearly
constitutes a remarkable robustness enhancement in comparison to
$(1-p)^N$, specially in the limit $\epsilon<<1$ [see Fig.
\ref{fig:entanglementinbipartitions} b)]. In addition, numerical
tests show that the approximation actually holds up to relatively
large $p$. For example, with $\epsilon=0.1$ and $N=20$, the
approximation by the exponential above is excellent up to
$p\approx0.2$. Furthermore, this in turn tends to
$\mathcal{N}\approx(1-p)^{N \epsilon}$ as  $N$ increases, an
exponential decay as that of \eqref{ghzstate} but with the
exponent damped by the factor $\epsilon$. In all cases, the less
the noise deviates from exact dephasing ($\alpha_{Z}=1$), the
slower is the decay of entanglement.

We emphasize that the enhancement is achieved without any overhead
in complex logical-qubit encodings, just through local Hadamard
rotations. Interestingly, these rotations correspond to a
qubit-basis transversal to the one privileged  by to the noise. In
particular, for GHZ entanglement, the robust qubit-basis is
exactly the one orthogonal to that defined by the only pure states
immune to the noise. Numerical optimizations up to $N=5$ show that
no other single-qubit basis yields slower decay of negativity than
that of \eqref{ghzstatetrans}, although other single-qubit bases
attain the same decay (up to numerical precision).

\emph{Quantum metrology with dephased resources.---} For mixed
states, higher entanglement does not necessarily imply better
performance at fulfilling some physical task. In what follows, we
show that our local-unitary protection enhances also the
robustness against local dephasing of GHZ states as resources for
phase estimation~\cite{metrology} (For a frequency estimation model see for instance \cite{Chaves2013}).

\begin{figure} [!t]
\centering
\includegraphics[width=0.8\linewidth]{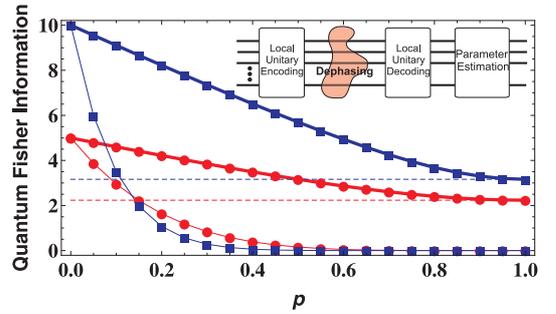}
\caption {(Color online) If, before local dephasing acts (inset),
GHZ states are encoded into transversal qubit-bases, and then
decoded before the parameter estimation, their usefulness for the
estimation is deteriorated exponentially less. The resulting
(square root of the) Fisher information
$\sqrt{\mathcal{F}^{PD}_{T}/4}$ (thick lines) is exponentially above
that of bare GHZ states, $\sqrt{\mathcal{F}^{PD}/4}$ (thin lines).
Blue squares correspond to $N=10$ qubits, whereas red circles to
$N=5$. Dashed thin lines delimit the maxima of
$\sqrt{\mathcal{F}/4}$ over the separable states. Notice that while
$\mathcal{F}^{PD}_{T}$ lies above the classical limit
throughout the dynamics, $\mathcal{F}^{PD}$ crosses it
exponentially fast.} \label{fig:Fisherinf}
\end{figure}


We consider a Hamiltonian $\hat{H}=\lambda{\sum_{k=1}^{N}} Z_{k}$,
with unknown parameter $\lambda$.  The associated unitary
evolution operator over a time $t$ is $U_{\phi }=e^{-it\hat{H}}$,
with $\phi\doteq\lambda t$  the phase to be estimated. A generic
$N$-qubit probe state $\rho$ accordingly transforms as
$\rho\rightarrow \rho_{\phi}\doteq U_{\phi }\rho U^{\dagger}_{\phi
}$. The statistical deviation $\delta\phi$ in the estimation can
be bounded  as $\delta\phi\geq1/\left(2\sqrt{\nu\mathcal{F}(
\rho_{\phi})}\right)$~\cite{CramerRao}. Here, $\nu$ represents the
number of runs of the estimation, and $\mathcal{F}( \rho_{\phi})$
is the quantum Fisher Information \cite{fisher} of $\rho_{\phi}$, which measures how much information about $\phi$ can be extracted
from $\rho_{\phi}$. $\mathcal{F}$ can be compute as 
\begin{equation}
\mathcal{F}= {\displaystyle\sum\limits_{k,p_{k}\neq0}}
\frac{\left(  p\left( \phi\right)  _{k}^{\prime}\right)
^{2}}{p\left( \phi\right) _{k}}+ {\displaystyle\sum_{j<k}}
\frac{4\left(  p_{j}-p_{k}\right) ^{2}}{p_{j}+p_{k}}\left\vert
\left\langle \omega_{j}^{\prime}\left( \phi\right)  \right.
\left\vert \omega_{k}\left( \phi\right) \right\rangle \right\vert
^{2}, \label{fishermixed}
\end{equation}
$p\left(  \phi\right)  _{k}$ and $\left\vert \omega_{j}\left(
\phi\right) \right\rangle $ being respectively the eigenvalues and
eigenvectors of the density matrix $\rho_{\phi}$ and
$p\left( \phi\right) _{k}^{\prime}=dp_{k}/d\phi$ the
derivative in relation to the parameter $\phi$. If $\rho_{\phi}$ is separable, $\mathcal{F}(
\rho_{\phi})$ is always limited as $\mathcal{F}(\rho_{\phi})\leq
4N$. However, entangled states can reach the maximal value
$\mathcal{F}_{\text{max}}=4N^{2}$, which yields a quadratic gain in
precision. This is the maximal gain compatible with the
uncertainty principle and is therefore known as the Heisenberg
limit.

The Fisher information of locally dephased GHZ state
$\rho=\Lambda^{PD}(\ket{{{\Phi_+}^{N}}})$ is $\mathcal{F}^{PD}=
4N^2(1-p)^{2N}$. Clearly, the quantum gain is obliterated by
decoherence exponentially fast. However, if before dephasing takes
place one locally rotates the resource GHZ state to
\eqref{ghzstatetrans}, and then, before the estimation, undoes the
rotation, the resulting Fisher information
$\mathcal{F}^{PD}_{T}\doteq \mathcal{F}\big({H^{\dagger}}^{\otimes
N}\Lambda^{PD}(\ket{{{\Phi_+}_T^{N}}})H^{\otimes N}\big)$ is
\begin{equation}
\mathcal{F}^{PD}_{T}= 4N^{2}(1-p)^{2}+16N\big(
1-\frac{p}{2}\big)\frac{p}{2}.
\end{equation}
The robustness-enhancement is thus exponential also for the
accuracy in phase estimations (see Fig.~\ref{fig:Fisherinf}).
\begin{figure} [!t]
\centering
\includegraphics[width=0.8\linewidth]{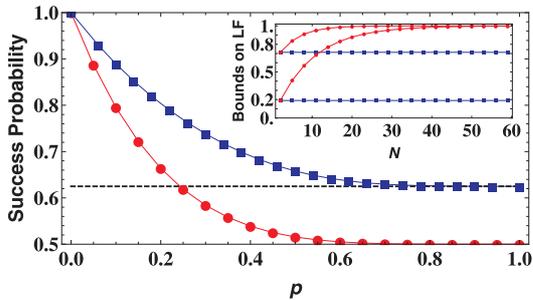}
\caption {(Color online) Probabilities of success for the
Mermin-Klyshko CCP as a function of $p$ and for $N=5$ qubits. Blue
squares correspond to the correlations in
$\Lambda^{PD}(\ket{{{\Phi_+}_T^{N}}})$, whereas red circles to
those in $\Lambda^{PD}(\ket{{{\Phi_+}^{N}}})$. The thin dashed
line in turns delimits the maximum classical probability of
success. With the local-unitary protection there is a quantum gain
throughout the noisy dynamics, while the dephased bare-GHZ
correlations enter the classical domain exponentially fast (not
shown). This behavior is also reflected by the local fraction LF
of the correlations. The inset shows lower and upper bounds of LF,
as a function of $N$ and for $p=0.1$, both for
$\Lambda^{PD}(\ket{{{\Phi_+}_T^{N}}})$ (blue squares) and for
$\Lambda^{PD}(\ket{{{\Phi_+}^{N}}})$ (red circles). While the
bounds for the latter tend to unit exponentially fast, those for
the former are size-independent.} \label{fig:CCP}
\end{figure}
\par\emph{Nonlocal computations with dephased resources.---}
Finally, we show that our local-unitary protection scheme enhances
the robustness against local dephasing of the nonlocality of
quantum correlations. Specifically, we focus on the performance of
dephased GHZ states to assist in solving distributed-computing
tasks, also known as communication-complexity problems (CCPs)
\cite{CCP,CCPBrukner}.

In the considered CCPs, $N$ distant users, assisted by some
correlations and a restricted amount of communication, must
locally calculate the value of a given function $f$. For every
Bell inequality there exists a CCP that can be solved with a higher
probability of success with nonlocal correlations than with any
classical resource if, and only if, the correlations violate the
inequality \cite{CCPBrukner}. The probability of success in the
CCP is $P_{S}= (1/2)(1+\beta_{Q}/\beta_{NL})$, where $\beta_{Q}$
is the Bell violation by the nonlocal correlations in the resource
quantum state and $\beta_{NL}$ the  maximum violation over
arbitrary nonlocal correlations.

As an example, we consider  the  CCP associated with the
Mermin-Klyshko (MK) inequality for $N$-bit correlations~\cite{MK}.
We obtain that $\Lambda^{PD}(\ket{{{\Phi_+}_T^{N}}})$ violates the
inequality for all $0\leq p< 1$, and its violation is bigger than
that by $\Lambda^{PD}(\ket{{{\Phi_+}^{N}}})$. This leads to the
enhancements of $P_{S}$ as the one plotted in Fig.~\ref{fig:CCP}.
Also, in the inset of Fig.~\ref{fig:CCP}, we have plotted the
dependence with $N$ of simple lower and upper bounds of the local
fraction of the correlations, which quantifies the fraction of
events describable by a local model \cite{epr2}. The bounds were
calculated as explained in Ref. \cite{chaves}. From these, one can see
that while, for large $N$, the local fraction of dephased bare GHZ
states tends to unit exponentially fast, that of  transversal GHZ
states stays always below a constant value ($\approx 0.7$).

\emph{Conclusions.---} There are many relevant situations in which
multi-particle entanglement is subject to local noise, \eg the
distribution of entangled particles to many distant parties or the
storage of these particles into different quantum memories. Here
we have focused on the physically relevant case in which the noise
has a privileged direction and  have provided a simple and
experimentally friendly recipe to enhance the robustness of
quantum correlations. We have shown that a simple local change of
bases, while preserving the correlation properties of the state,
significantly improves its robustness. For general graph states,
we have derived  bounds on the decay entanglement and relative
entropy of quantumness that are independent of $N$. In the case of
GHZ states, we have shown not only that the local-unitary encoding
neutralizes their exponential decay with the system size, but also
that an exponential improvement is still observed when there are
deviations from the ideal case. In addition, the robustness of the
usefulness of GHZ states as resources for parameter estimation and
nonlocal computations is equivalently enhanced.

The enhancement introduces no cost at all in extra particles. We
believe that the fact that an exponential enhancement is achieved
through such an extremely simple scheme, makes the present
passive-protection approach highly relevant to many current
experimental platforms.

\emph{Acknowledgements.---} This work was supported by the
European ERC Starting grant PERCENT and the Q-Essence project, the
Spanish FIS2010-14830 project and a Juan de la Cierva grant, and
Caixa Catalunya.


\appendix
\section{Robustness law for the relative entropy of
quantumness} \label{relativeent} The relative entropy of
quantumness is a discord-like measure of quantum correlations (see
\cite{Kavan} and references therein). That is, it encapsulates
entanglement but does not restrict to it. Like all variants of
discord, it is not non-increasing under local operations assisted
by classical communication (LOCC). Indeed, it is not even
non-increasing under general local operations \cite{Streltsov}.
Still, in this appendix we show that a robustness law equivalent
to (4) applies to it too.

\par The relative entropy of quantumness $Q_S(\rho)$ of an $N$-qubit state $\rho$ is defined \cite{Kavan}
as $Q_S(\rho)\doteq S(\rho||\xi_{\text{min}})$. The von Neumann
relative entropy $S(\rho||\xi_{\text{min}})\doteq
-\text{Tr}[\rho\log(\xi_{\text{min}})]+\text{Tr}[\rho\log(\rho)]$
between states $\rho$ and $\xi_{\text{min}}$ measures how
distinguishable they are. State $\xi_{\text{min}}$ is the closest
classical state to $\rho$, in the sense of minimizing the relative
entropy with $\rho$ over all exclusively-classically correlated
$N$-qubit states $\xi\doteq\sum_{i_1 \hdots i_N}p_{i_1 \hdots
i_N}\ket{i_1 \hdots i_N}\bra{i_1 \hdots i_N}$, with $p_{i_1 \hdots
i_N}$ any probability distribution and $\{\ket{i_1 \hdots i_N}\}$
any $N$-qubit basis. We show in what follows that
\begin{eqnarray}
\label{orderingdiscord}
Q_S\big(\Lambda^{PD}(\ket{{{\Phi_+}^{N}_{T}}})\big)\geq\hdots\geq
Q_S\big(\Lambda^{PD}(\ket{{{\Phi_+}^{2}_{T}}})\big),
\end{eqnarray}
for all $N\geq2$.
\par As in the derivation of \eqref{ordering}, we consider a single-qubit $Z$ measurement acting on $\Lambda^{PD}(\ket{{\Phi_+}^{N}_{T}})$, which leaves the system in state $\frac{1}{2}\big(\ket{0}\bra{0}\otimes{\rho_+}_{T}^{N-1}+\ket{1}\bra{1}\otimes{\rho_-}_{T}^{N-1}\big)$.
As said, $Q_S$ cannot be guaranteed not to increase under generic
local maps. However, for the particular case when the local maps
are unital (those mapping the identity operator $\openone$ into
itself), it was shown in Ref. \cite{Streltsov} that $Q_S$ is
non-increasing. Thus, since a single-qubit $Z$ measurement is a
local unital map, we have that
\begin{eqnarray}
\label{orderingQ} \nonumber
&&Q_S\big(\Lambda^{PD}(\ket{{{\Phi_+}^{N}_{T}}})\big)\geq\\
&&
Q_S\Big(\frac{1}{2}\big(\ket{0}\bra{0}\otimes{\rho_+}_{T}^{N-1}+\ket{1}\bra{1}\otimes{\rho_-}_{T}^{N-1}\big)\Big).
\end{eqnarray}
Besides, by definition it is $Q_S\Big(\frac{1}{2}\big(\ket{0}\bra{0}\otimes{\rho_+}_{T}^{N-1}+\ket{1}\bra{1}\otimes{\rho_-}_{T}^{N-1}\big)\Big)\doteq S\Big(\frac{1}{2}\big(\ket{0}\bra{0}\otimes{\rho_+}_{T}^{N-1}+\ket{1}\bra{1}\otimes{\rho_-}_{T}^{N-1}\big)\big|\big|\tilde{\xi}_{\text{min}}^N\Big)$, with $\tilde{\xi}_{\text{min}}^N
$ the closest classical state to
$\frac{1}{2}\big(\ket{0}\bra{0}\otimes{\rho_+}_{T}^{N-1}+\ket{1}\bra{1}\otimes{\rho_-}_{T}^{N-1}\big)$.
Next,  using the definition of the relative entropy in terms of
traces, taking the partial trace over the measured qubit, and
after a straightforward calculation, we obtain
\begin{eqnarray}
\label{Slog} \nonumber
&&S\Big(\frac{1}{2}\big(\ket{0}\bra{0}\otimes{\rho_+}_{T}^{N-1}+\ket{1}\bra{1}\otimes{\rho_-}_{T}^{N-1}\big)\big|\big|\tilde{\xi}_{\text{min}}^N\Big)=\\
\nonumber
&&\frac{1}{2}\Big[S\Big({\rho_+}_{T}^{N-1}\Big|\Big|\frac{\bra{0}\tilde{\xi}_{\text{min}}^N\ket{0}}{\text{Tr}[\bra{0}\tilde{\xi}_{\text{min}}^N\ket{0}]}\Big)\\
\nonumber
&&+S\Big({\rho_-}_{T}^{N-1}\Big|\Big|\frac{\bra{1}\tilde{\xi}_{\text{min}}^N\ket{1}}{\text{Tr}[\bra{1}\tilde{\xi}_{\text{min}}^N\ket{1}]}\Big)\\
&&-\log\big(2x\big)-\log\big(2(1-x)\big)\Big],
\end{eqnarray}
with $x\doteq\text{Tr}[\bra{0}\tilde{\xi}_{\text{min}}^N\ket{0}]$
and
$1-x\doteq\text{Tr}[\bra{1}\tilde{\xi}_{\text{min}}^N\ket{1}]$.
\par Now, from \eqref{Slog} we immediately obtain an explicit form for $\tilde{\xi}_{\text{min}}^N$: It must be
\begin{eqnarray}
\tilde{\xi}_{\text{min}}^N=\frac{1}{2}\big(\ket{0}\bra{0}\otimes{\xi_{\text{min}}}_+^{N-1}+\ket{1}\bra{1}\otimes{\xi_{\text{min}}}_-^{N-1}\big),
\end{eqnarray}
 with ${\xi_{\text{min}}}_+^{N-1}$ and ${\xi_{\text{min}}}_-^{N-1}$ the closest classical $(N-1)$-qubit states to ${\rho_+}_{T}^{N-1}$ and ${\rho_-}_{T}^{N-1}$, respectively. This is due to the following observations: ({\it i}) Clearly, with this form, both the first and second lines after the equality in \eqref{Slog} are minimized. ({\it ii}) The minimum of $-\log\big(2x\big)-\log\big(2(1-x)\big)$ over $x\in[0,1]$ is zero, which is precisely the value the form of $\tilde{\xi}_{\text{min}}^N$ above yields. This leads us to
\begin{eqnarray}
\label{QS} \nonumber
&&Q_S\Big(\frac{1}{2}\big(\ket{0}\bra{0}\otimes{\rho_+}_{T}^{N-1}+\ket{1}\bra{1}\otimes{\rho_-}_{T}^{N-1}\big)\Big)=\\
\nonumber
&&\frac{1}{2}\Big[S\big({\rho_+}_{T}^{N-1}\big|\big|{\xi_{\text{min}}}_+^{N-1}\Big)+S\Big({\rho_-}_{T}^{N-1}\big|\big|{\xi_{\text{min}}}_-^{N-1}\Big)\Big]\\
&&\doteq\frac{1}{2}\Big(Q_S\big({\rho_+}_{T}^{N-1}\big)+Q_S\big({\rho_-}_{T}^{N-1}\big)\Big),
\end{eqnarray}
but, since states
${\rho_+}_{T}^{N-1}\doteq\Lambda^{PD}(\ket{{{\Phi_+}^{N-1}_{T}}})$
and
${\rho_-}_{T}^{N-1}\doteq\Lambda^{PD}(\ket{{{\Phi_-}^{N-1}_{T}}})$
are local-unitarily equivalent, they posses exactly the same
amount and type of quantum correlations. Therefore, the last line
of \eqref{QS} equals
$Q_S\big(\Lambda^{PD}(\ket{{{\Phi_+}^{N-1}_{T}}})\big)$ and,
together with \eqref{orderingQ}, renders
$Q_S\big(\Lambda^{PD}(\ket{{{\Phi_+}^{N}_{T}}})\big)\geq
Q_S\big(\Lambda^{PD}(\ket{{{\Phi_+}^{N-1}_{T}}})\big)$. Again as
in the derivation of (4), iterating this reasoning $N-2$ times one
arrives at \eqref{orderingdiscord}. $\square$

\section{Robustness law for generic (possibly mixed) graph states}
Here, we extend bounds (4) and \eqref{orderingdiscord} first to
arbitrary pure graph states, and then to globally-depolarized
arbitrary graph states. To encompass both bounds with the same
notation, we use in what follows $C$ to denote a generic measure
of quantum correlations, wich can either be an entanglement
monotone, $C=E$, or the relative entropy of quantumness, $C=Q_S$,
defined in App. A.

\par For every qubit of any connected $N$-qubit graph state
$\ket{\mathcal{G}^N}$, a measurement in either the $Z$ or the $X$
bases leaves the remaining qubits in a connected $(N-1)$-qubit
graph state $\ket{\mathcal{G}^{N-1}}$ (or in a state
local-unitarily equivalent to it, depending on the measurrment
outcome)~\cite{graph_review}. One can thus apply the the same
machinery used in the derivations of (4) and
\eqref{orderingdiscord} and arrive at the size-independent bound
\begin{eqnarray}
\label{orderinggraphs}
C\big(\Lambda^{PD}(\ket{\mathcal{G}_T^N})\big)\geq\hdots\geq
C\big(\Lambda^{PD}(\ket{\mathcal{G}_T^2})\big),
\end{eqnarray}
for all $N\geq2$. Here, $\ket{\mathcal{G}_T^2}$ is a two-qubit
graph state (local-unitarily equivalent to
$\ket{{\Phi_+}^{2}_{T}}$), and $\ket{\mathcal{G}_T^N}$ is obtained
by applying single-qubit Hadamard rotations to some of the qubits
in $\ket{\mathcal{G}^N}$ (those corresponding to the
above-mentioned $X$ measurements). Finally, the same arguments
hold even for imperfect initial states of the form
$v\ket{\mathcal{G}^N}\bra{\mathcal{G}^N}+(1-v)\frac{\openone}{2^N}$,
where $0\leq v\leq1$ is some visibility. The resulting robustness
law is then
\begin{eqnarray}
\label{orderinggraphsmixed} \nonumber
C\Big(\Lambda^{PD}\big(v\ket{\mathcal{G}_T^N}\bra{\mathcal{G}_T^N}+(1-v)\frac{\openone}{2^N}\big)\Big)\geq\hdots\\
\geq
C\Big(\Lambda^{PD}\big(v\ket{\mathcal{G}_T^2}\bra{\mathcal{G}_T^2}+(1-v)\frac{\openone}{4}\big)\Big).
\end{eqnarray}
\section{Asymptotic value of negativity under exact dephasing}
\label{Asymptotic} Here, we first show that negativity (5) reduces
to (6) when $\alpha_{Z}=1$, and then that the latter tends to the
$N$-independent value $1-p$ in the limit $N\to\infty$, for all
$0\leq p<1$.

\par First, taking $\alpha_{Z}=1$ and $\alpha_{X}=0=\alpha_{Y}$ in (5) leads, through a simple and straightforward calculation, to
\begin{eqnarray}
\label{negZ2} \nonumber
\mathcal{N}_{(\alpha_{Z}=1)}&=&(1-p)\sum_{\mu=0}^{\lfloor\frac{N'}{2}\rfloor} \binom{N'}{\mu}\bigg|\frac{p}{2}^{\mu}(1-\frac{p}{2})^{N'-\mu}\\
&-&\frac{p}{2}^{N'-\mu}(1-\frac{p}{2})^{\mu}\bigg|,
\end{eqnarray}
with $N'\doteq N-1$. Let us see that the absolute value inside
this summation can be removed. Notice that, for all $0\leq p\leq
1$, and for any $0\leq\mu\leq\lfloor\frac{N'}{2}\rfloor$, it is
\begin{eqnarray}
\label{absvalue} \nonumber
1-\frac{p}{2}&\geq& \frac{p}{2}\Rightarrow\\
\nonumber
(1-\frac{p}{2})^{N'-2\mu}&\geq& \frac{p}{2}^{N'-2\mu}\Rightarrow\\
\nonumber
\frac{p}{2}^{\mu}(1-\frac{p}{2})^{N'-\mu}&\geq&\frac{p}{2}^{N'-\mu}(1-\frac{p}{2})^{\mu}.
\end{eqnarray}
Therefore
$\big|\frac{p}{2}^{\mu}(1-\frac{p}{2})^{N'-\mu}-\frac{p}{2}^{N'-\mu}(1-\frac{p}{2})^{\mu}\big|\equiv\frac{p}{2}^{\mu}(1-\frac{p}{2})^{N'-\mu}-\frac{p}{2}^{N'-\mu}(1-\frac{p}{2})^{\mu}$,
what implies
\begin{eqnarray}
\label{negZ3} \nonumber
\mathcal{N}_{(\alpha_{Z}=1)}&=&(1-p)\sum_{\mu=0}^{\lfloor\frac{N'}{2}\rfloor} \binom{N'}{\mu}\bigg[\frac{p}{2}^{\mu}(1-\frac{p}{2})^{N'-\mu}\\
&-&\frac{p}{2}^{N'-\mu}(1-\frac{p}{2})^{\mu}\bigg].
\end{eqnarray}
\par Next we show that $\lim\limits_{N\to\infty}\mathcal{N}_{(\alpha_{Z}=1)}=1-p$, for all $0\leq p<1$. To this end, see first that
\begin{eqnarray}
\label{sum} \nonumber
\sum_{\mu=0}^{\lfloor\frac{N'}{2}\rfloor} \binom{N'}{\mu}\frac{p}{2}^{N'-\mu}(1-\frac{p}{2})^{\mu}\equiv&&\\
\sum_{\mu=\tilde{N}}^{N'}\binom{N'}{\mu}\frac{p}{2}^{\mu}(1-\frac{p}{2})^{N'-\mu},&&
\end{eqnarray}
with
\begin{displaymath}
\tilde{N}\doteq\left\{
\begin{array}{ll}
\lfloor\frac{N'}{2}\rfloor+1
 & \textrm{if $N$ is even, }\\
\lfloor\frac{N'}{2}\rfloor
 & \textrm{if $N$ is odd.}
\end{array} \right.
\end{displaymath}
Using  \eqref{sum}, one rewrites negativity (\ref{negZ})  as
$\mathcal{N}_{(\alpha_{Z}=1)}=(1-p)(S_1-S_2)$, where $S_1$ and
$S_2$ are the following sums:
\begin{subequations}
\begin{align}
S_1\doteq\sum_{\mu=0}^{\lfloor\frac{N'}{2}\rfloor} \binom{N'}{\mu}\frac{p}{2}^{\mu}(1-\frac{p}{2})^{N'-\mu},\\
\label{S2}
S_2\doteq\sum_{\mu=\tilde{N}}^{N'}\binom{N'}{\mu}\frac{p}{2}^{\mu}(1-\frac{p}{2})^{N'-\mu}.
\end{align}
\end{subequations}

\par Now, invoking the binomial theorem, we notice that
\begin{displaymath}
S_1+S_2\equiv\left\{
\begin{array}{ll}
1 & \textrm{if $N$ is even, }\\
1+\binom{N'}{N'/2}\frac{p}{2}^{\frac{N'}{2}}(1-\frac{p}{2})^{\frac{N'}{2}}
& \textrm{if $N$ is odd.}
\end{array} \right.
\end{displaymath}
Therefore, negativity ((\ref{negZ})) can be expressed as
$\mathcal{N}_{(\alpha_{Z}=1)}=(1-p)(1-2S_2)$, when $N$ is even,
and as
$\mathcal{N}_{(\alpha_{Z}=1)}=(1-p)\Big(1-2S_2+\binom{N'}{N'/2}\frac{p}{2}^{\frac{N'}{2}}(1-\frac{p}{2})^{\frac{N'}{2}}\Big)$,
when $N$ is odd.

\par Finally, we show first that $\binom{N'}{N'/2}\frac{p}{2}^{\frac{N'}{2}}(1-\frac{p}{2})^{\frac{N'}{2}}\to0$ as $N'\to\infty$, and then that also $S_2\to0$ as $N'\to\infty$, which finishes the proof. For sufficiently large $N'$, we can apply Stirling's approximation for the factorial: $N'!\to\sqrt{2\pi N'}{\frac{N'}{e}}^{N'}$. So, we obtain that, for $N'\to\infty$,
\begin{eqnarray}
\nonumber
\binom{N'}{N'/2}\frac{p}{2}^{\frac{N'}{2}}(1-\frac{p}{2})^{\frac{N'}{2}}&\to&\\
\nonumber
\sqrt{\frac{2}{\pi N'}}2^{N'}\frac{p}{2}^{\frac{N'}{2}}(1-\frac{p}{2})^{\frac{N'}{2}}&\doteq&\\
\nonumber
\sqrt{\frac{2}{\pi}\frac{\gamma^{N'}}{N'}}&\to&\\
\nonumber 0,&&
\end{eqnarray}
where the convergence of the last limit is guaranteed by the fact
that $0\leq\gamma\doteq2^2\frac{p}{2}(1-\frac{p}{2})<1$, for all
$0\leq p\leq 1$.

\par Now, from definition \eqref{S2} and the facts that $\binom{N'}{\mu}\leq\binom{N'}{N'/2}$ and $\frac{p}{2}^{\mu}(1-\frac{p}{2})^{N'-\mu}\leq\frac{p}{2}^{\frac{N'}{2}}(1-\frac{p}{2})^{\frac{N'}{2}}$, for all $\tilde{N}\leq\mu\leq N'$, we see that sum $S_2$ is bounded from above as
\begin{displaymath}
S_2\leq\left\{
\begin{array}{ll}
\frac{N}{2}\binom{N'}{N'/2}\frac{p}{2}^{\frac{N'}{2}}(1-\frac{p}{2})^{\frac{N'}{2}} & \textrm{if $N$ is even, }\\
\big(\frac{N}{2}+1\big)\binom{N'}{N'/2}\frac{p}{2}^{\frac{N'}{2}}(1-\frac{p}{2})^{\frac{N'}{2}}
& \textrm{if $N$ is odd.}
\end{array} \right.
\end{displaymath}
Invoking once more Stirling's approximation, using $N'= N-1$ and
the definition of $\gamma$ above, we have that for large $N$ these
bounds are approximately given by
\begin{displaymath}
S_2\leq\left\{
\begin{array}{ll}
\sqrt{\frac{1}{2\pi}\frac{N^2}{N-1}\gamma^{N-1}} & \textrm{if $N$ is even, }\\
\sqrt{\frac{1}{2\pi}\frac{(N+2)^2}{N-1}\gamma^{N-1}} & \textrm{if
$N$ is odd.}
\end{array} \right.
\end{displaymath}
As $N\to\infty$, both bounds tend to the quantity
$\sqrt{\frac{1}{2\pi}N\gamma^{N-1}}$, whose limiting value is 0
again because $0\leq\gamma<1$. $\square$

\end{document}